\documentclass[
 reprint,
 amsmath,amssymb,
 aps,
]{revtex4-2}

\usepackage{graphicx}
\usepackage{gensymb}\usepackage{braket}
\usepackage[qm]{qcircuit}
\usepackage[dvipsnames]{xcolor}
\usepackage{float}
\usepackage{booktabs,siunitx}
\usepackage{dcolumn}
\usepackage{bm}
\usepackage{subfigure}

\newcommand{\comment}[1]{}

\newcommand{\rb}{\textbf{r}}
\newcommand{\rbb}{\textbf{r'}}
\renewcommand{\sp}{\hspace{0.1cm}}
\usepackage{ulem}

\begin{document}

\preprint{APS/123-QED}

\title{Integrable atomtronic interferometry
}

\author{D.S. Gr\"un$^1$, L.H. Ymai$^2$, K. Wittmann W.$^1$, A.P. Tonel$^2$, A. Foerster$^1$, J. Links$^3$}

\affiliation{$^1$ Instituto de F\'{i}sica da UFRGS, Avenida Bento Gon\c{c}alves 9500, Porto Alegre, Rio Grande do Sul, Brazil,}
\affiliation{$^2$ Universidade Federal do Pampa, Av. Maria
Anuncia\c{c}\~ao Gomes de Godoy 1650, Bag\'e, Rio Grande do Sul, Brazil,}
\affiliation{$^3$ School of Mathematics and Physics, The University of Queensland,
Brisbane, Queensland, Australia.}

\date{\today}

\begin{abstract}
High sensitivity quantum interferometry requires more than just access to entangled states.  It is achieved through deep understanding of quantum correlations in a system. Integrable models offer the framework to develop this understanding. 
We communicate the design of interferometric protocols for an integrable model that describes the interaction of bosons in a four-site configuration. Analytic formulae for the quantum dynamics of certain observables are computed. These expose the system's functionality as both an interferometric identifier, and producer, of NOON states. 
Being equivalent to a controlled-phase gate acting on two hybrid qudits, this system also highlights an equivalence between Heisenberg-limited interferometry and quantum information. These results are expected to open new avenues for integrability-enhanced atomtronic technologies.

\end{abstract}

\maketitle

\textit{Introduction.}-- Recent developments in the manipulation of wave-like properties in matter are driving a raft of atom-interferometric applications, in the vicinity of the Heisenberg limit,  within the field of quantum metrology  \cite{bongs,pezze}.  
It has long been recognized that  the ability to effectively and efficiently harness quantum interference is equivalent to  implementing certain tasks in quantum computation \cite{rosetta}. Nowadays, ultracold quantum gases are proving to be successful in enabling quantum simulations, for phenomena such as  quantum magnetism and topological states of matter, beyond the capabilities of classical supercomputers \cite{gross}. Through a confluence of these types of  investigations, there are several efforts to push research towards  designs for atomtronic devices \cite{pepino,boshier,amico}, based on circuits with atomic currents \cite{nunnenkamp,stiebler,ragole,haug1}. These devices promise high levels of control in the manipulation of  many-body systems, leading to advanced sensitivity in metrology  \cite{pand} and other quantum technologies \cite{olsen,karin,haug,polo,comp}.  

Around a decade ago \cite{santos} a class of models was identified for physical realization of an interferometer, using dipolar atoms. The Hamiltonian governing the time evolution of the system is a generalized Bose-Hubbard model on four sites, with closed boundary conditions and long-ranged interactions.    We begin by pinpointing a set of {\it integrable} couplings for the Hamiltonian. That is, choices of parameters for which there are four conserved operators, equal to the number of degrees of freedom.  The property of integrability has two significant impacts: (i) integrable systems have unique properties, such as Poisson distribution in energy level statistics \cite{poil}, absence of chaotic behaviors \cite{lea}, and non-standard thermal equilibration \cite{cala}. The quantum Newton cradle  \cite{newton} provided experimental verification of the latter; (ii) mathematically, integrability facilitates tractable, closed-form  formulae to describe the physics.

\begin{figure}[!h]
    \centering
\includegraphics[width=8cm]{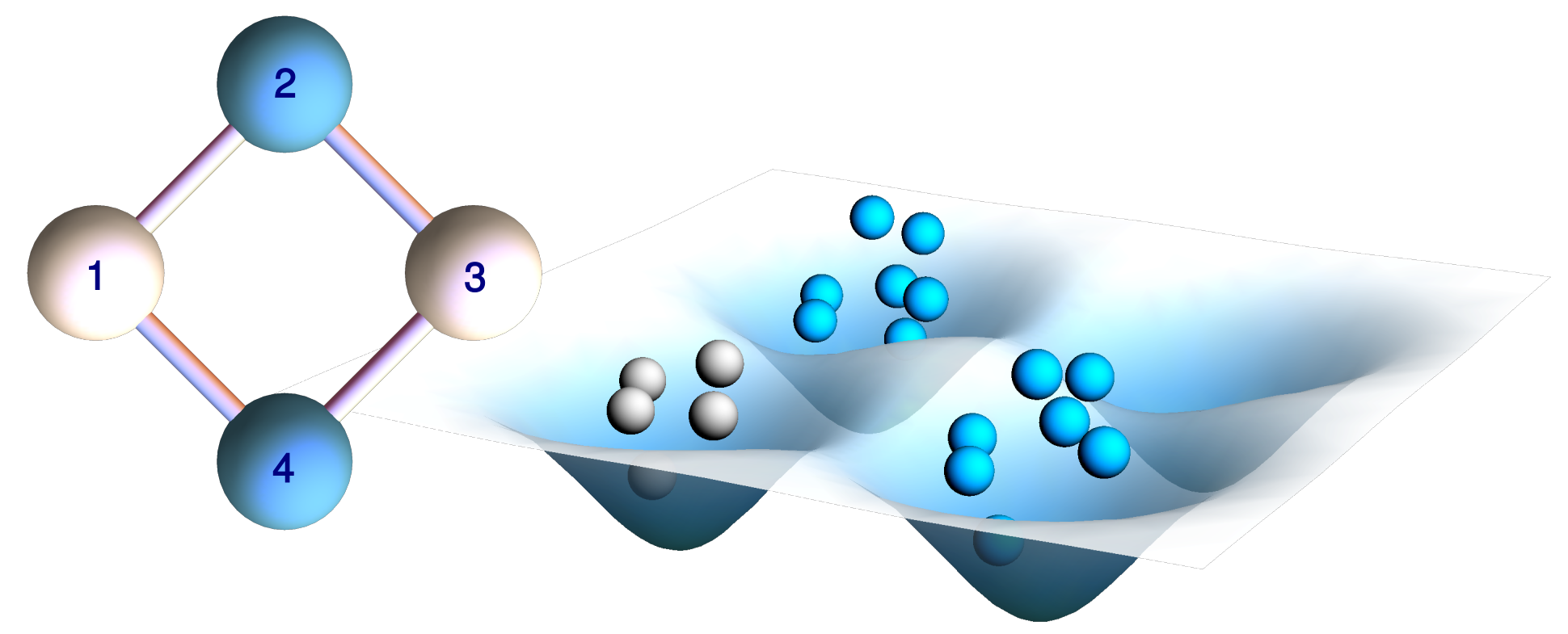}
    \caption{Schematic representation of the interferometric circuit with tunneling between nearest neighbors. An initial state is prepared with $M$ particles in site 1, and $P$ particles in a (generally entangled) state across sites 2 and 4. 
    After Hamiltonian time-evolution, measurement of particle number at site 3 is used to deduce information about the initial, or post-measurement, state across sites 2 and 4.}
    \label{fig:one}
\end{figure}

In our study we utilize the conserved operators of the integrable system to guide the design of measurement protocols for interferometric tasks (see Fig. \ref{fig:one}). Our results are applicable in a particular regime, designated as {\it resonant tunneling}, whereby the energy levels separate into distinct bands. Through an effective Hamiltonian approach, the entire energy spectrum and structure of eigenstates becomes explicit for resonant tunneling. Moreover, the system's behavior is clear in quantum information theoretic terms. The interferometer is equivalent to a system of two {\it hybrid qudits} \cite{sanders}, and the time-evolution of states is equivalent to the operation of a {\it controlled-phase} gate \cite{muthu,bren}. We describe proof of principle examples of high-fidelity measurement protocols to identify and produce certain NOON states \cite{pezze,rosetta,pryde,mitchell,resch}, 
We also provide a physical-feasibility analysis of the system with first-principles calculations of the Hamiltonian parameters within an explicit Bose-atom setup (see Supplemental Material A).    

\textit{The model.}-- An extended Bose-Hubbard Hamiltonian on a square plaquette has the form \cite{bar,lah}

\begin{flalign}
\begin{split}
    H &= \frac{U_0}{2}\sum_{i=1}^{4}N_i(N_i - 1) + \sum_{i=1}^{4}\sum_{j=1,j\neq i}^{4}\frac{U_{ij}}{2} N_i N_j  \\
    &\quad - \frac{J}{2}[(a_1^\dagger + a_3^\dagger)(a_2 + a_4) + (a_2^\dagger + a_4^\dagger)(a_1 + a_3)]. 
    \label{ham}
\end{split}
\end{flalign}
where $\{a_j,\,a_j^\dagger:j=1,2,3,4\}$ are canonical boson annihilation and creation operators, 
$U_0$ characterizes the short-range interactions between bosons at the same site, $U_{ij}=U_{ji}$ accounts for long-range (e.g. dipole-dipole) interactions between sites, and $J$ represents the tunneling strength between neighboring sites. The Hamiltonian commutes with the total particle number $N=N_1+N_2+N_3+N_4$ where $N_j=a_j^\dagger a_j$. Moreover, the Hamiltonian is integrable when $U_{13}=U_{24}=U_0$ and 
$U_{12}=U_{14}=U_{23}=U_{34}$. It acquires two additional conserved operators
\begin{flalign*}
&Q_1 = \frac{1}{2}(N_1 + N_3 - a_1^\dagger a_3 - a_3^\dagger a_1),\\
&Q_2 = \frac{1}{2}(N_2 + N_4 - a_2^\dagger a_4 - a_4^\dagger a_2),
\end{flalign*}
such that $[Q_1,Q_2] = [Q_{j},H] = [Q_{j},N] = 0$, $j=1,2$. Integrability results from derivation of the model through the Quantum Inverse Scattering Method. It is intimately related to exact solvability, due to the algebraic Bethe Ansatz \cite{tyfl}. Hereafter we only consider the integrable case.

\textit{Resonant tunneling regime.}-- It is straightforward to check that there are large energy degeneracies when $J=0$. From numerical diagonalization of (\ref{ham}), with $N$ particles and sufficiently small value of $J$, it is seen that the low-energy levels coalesce  into well-defined bands \cite{gwylf},  similar to that observed in an analogous integrable three-site model \cite{karin,arlei}.  In this regime, an effective Hamiltonian $H_{\rm eff}$ is obtained through consideration of second-order tunneling processes.     
For an initial Fock state $|M-l,P-k,l,k \rangle$, with total boson number $N=M+P$,
the effective Hamiltonian is a simple function of the conserved operators
\begin{equation}
    \label{hameff}
    H_{{\rm eff}} = (N+1)\Omega (Q_1 +  Q_2) -2 \Omega Q_1 Q_2,
\end{equation}
where  
$\Omega=J^2/(4U((M-P)^2-1))$ with $U=(U_{12}-U_0)/4$. This result is valid for $J\ll U(M-P)$, which characterizes the resonant tunneling regime. For time evolution under $H_{\rm eff}$, both $N_1+N_3=M$ and $N_2+N_4=P$ are constant. The respective $(M+1)$-dimensional subspace associated with sites 1 and 3 and $(P+1)$-dimensional subspace associated with sites 2 and 4 serve as two, coupled, hybrid qudits \cite{sanders}, and provide the state space for the relevant energy band. 
This yields a robust approximation for the dynamics under $(\ref{ham})$, which we benchmark below.  For later use we will designate the qudit associated with sites 1 and 3 as {\it qudit A}, and that associated with sites 2 and 4 as {\it qudit B}.

\begin{figure}[!h]
    \centering
        \includegraphics[width=8.6cm]{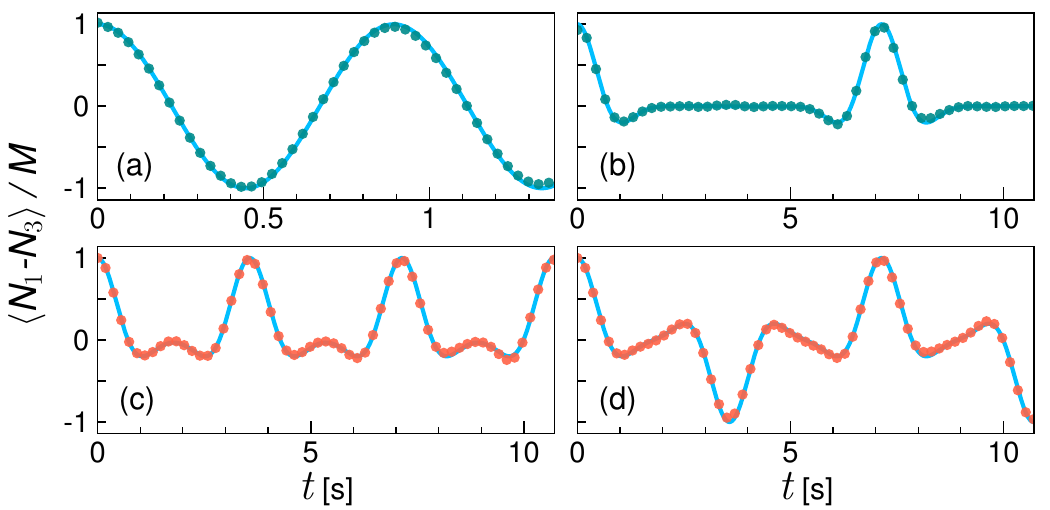}
    \caption{Time evolution of expected fractional imbalance $\braket{N_1-N_3}/M$ (dot points) for the  Hamiltonian (\ref{ham}) as a function of time $t$ in units of seconds, with 
    $U/J \simeq 1.2$, $U/\hbar \simeq 2\pi\times 19.5$ Hz, $J/\hbar \simeq 2\pi\times 16.2$ Hz, and  different initial states 
    :
    {\bf (a)} $\ket{4,0,0,0}$. {\bf (b)} $\ket{4,11,0,0}$. {\bf (c-d)} $(\ket{4,11,0,0}+\exp{(i\phi)}\ket{4,0,0,11})/\sqrt{2}$ with $\phi = 0$  {\bf (c)} and $\phi = \pi$  {\bf (d)}.
    The top panels display agreement with the formula (\ref{imb1}) (solid lines), while the bottom panels are in agreement with (\ref{imb2}) (solid lines).}
    \label{fig:two}
\end{figure}

It is found through Bogoliubov transformations that the spectrum of $H_{\rm eff}$ is $E_{{\rm eff}} = (N+1)\Omega (q_1 +  q_2) - 2 \Omega q_1 q_2$
with $q_1=0,...,M$ and $q_2=0,...,P$. Thus the time evolution under $H_{\rm eff}$ is recognized as a controlled-phase gate \cite{muthu,bren}. From here, several analytic results are accessible. For initial Fock state $|M,P,0,0\rangle$,  the expectation value of the fractional imbalance $\mathcal{I}(t)$ between sites 1 and 3 is (in units where $\hbar=1$)
\begin{align}
\mathcal{I}(t)  \equiv \langle N_1-N_3 \rangle/M
&=\cos((M+1) \Omega t)[\cos(\Omega t)]^{P}.
\label{imb1}
\end{align}
When $P=0$, there are harmonic oscillations in the imbalance. For non-zero $P$, the oscillations are no longer harmonic due to interference. For comparison, results from numerical diagonalization of (\ref{ham}) are shown in the upper panels of Fig. \ref{fig:two}

Other initial states can be studied, such as
\begin{align}
\ket{\Phi(\phi)} = \frac{1}{\sqrt{2}} \ket{M,P,0,0} + \frac{\exp{(i\phi)}}{\sqrt{2}}\ket{M,0,0,P},
\label{noon}
\end{align}
which is a product of a number state for site 1, vacuum for site 3 (qudit A), and a phase-dependent NOON state \cite{pezze,rosetta} across sites 2 and 4 (qudit B). We find the following result for the fractional imbalance between sites 1 and 3:
\begin{align} \label{imb2}
\langle N_1-N_3\rangle/M 
&=\cos((M+1)\Omega t)\left[\cos(\Omega t)\right]^P \\
+&\cos(\phi)\cos((M+1)\Omega t + {\pi P}/{2})[\sin(\Omega t)]^P.
\nonumber 
\end{align}
This formula provides excellent agreement with numerical calculations using (\ref{ham}). Examples are provided, for choices $\phi=0$ and $\phi=\pi$,
in the lower panels of Fig. \ref{fig:two} using experimentally feasible parameters evaluated in  Supplemental Material A.

\textit{NOON state identification and production.}--
The above results are sufficient to demonstrate the efficacy of the system to perform certain interferometric tasks.  First consider a black box processor ${\mathbb P}$ that outputs one of two possible NOON  states, either symmetric or antisymmetric. The output state, with particle number $P$, is loaded into qudit B. With $M$ particles in site 1 and zero in site 3 of qudit A, this composite initial state is given by  (\ref{noon}) with either $\phi=0$ (symmetric) or $\phi=\pi$ (anti-symmetric). Choose $M$ such that $N=M+P$ is odd, let the system   evolve for time $t_m=\pi/(2\Omega)$, and then measure the particle number at site 3.  According to (\ref{imb2}), there are only two possible measurement outcomes. One is to obtain the outcome  zero, which occurs with probability 1 when $\phi= \pi$. The other is to obtain the outcome $M$, which occurs with probability 1 when $\phi=0$ (cf. the lower panels of Fig. \ref{fig:two}, where the time of measurement is $t_m \simeq 3.57$s). Moreover, this measurement is non-destructive and the NOON state in qudit B is preserved \cite{footnote}. 

This analytic result is an excellent approximation for the behavior governed by (\ref{ham}). From numerical results using
the parameters of Fig. \ref{fig:two}, we find that the success probability when $\phi=0$ is $0.98334$, and it is $0.99383$ when $\phi=\pi$. This delivers a proof of principle example to show that the model (\ref{ham}) has capacity to perform interferometry with high accuracy. 

Remarkably, the earlier analysis on NOON state identification can now be inverted to show that the interferometer itself provides a high-fidelity simulation of the black box processor ${\mathbb P}$. For $\ket{\Psi_0} = \ket{M,P,0,0}$ with $N = M+P$ odd, it can be shown that  
\begin{align} 
\label{gennoon} \left|\Psi\left(t_m\right)\right>
&=\frac{(-1)^{(N+1)/2}}{2}\ket{M,P,0,0}+
\frac{1}{2}\ket{M,0,0,P}\\
&\quad +\frac{1}{2}\ket{0,P,M,0} + \frac{(-1)^{(N-1)/2}}{2} \ket{0,0,M,P}.
\nonumber
\end{align}
In accordance with the previous discussion, measurement at site 3 produces one of only two possible outcomes.  A measurement outcome of $M$ causes  wavefunction collapse such that the state of qudit B is the symmetric (antisymmetric) NOON state if $(N+1)/2$ is odd (even). Conversely, a measurement  outcome of zero causes wavefunction collapse with an antisymmetric (symmetric) NOON state in qudit B if $(N+1)/2$ is odd (even).

As before, it is useful to compare this result obtained from (\ref{hameff}) against the analogous predictions of (\ref{ham}). Numerically, using 
the parameters of Fig. \ref{fig:two}, we find that the outcome fidelity of this processor simulation for (\ref{ham}) is $0.97831$ for outcome zero, and $0.99298$ for outcome $M$, with respective probabilities of $0.49611$ and $0.47639$, close to the theoretically predicted values of 1/2 in each case. See Supplemental Material B for further details, including probabilities and fidelities for intermediate outcomes.

\textit{Entanglement and correlations}.-- The ability to produce NOON states as described above is clearly dependent on the ability to create entanglement. More important is the ability to create ``useful'' entanglement since, as emphasized in the review article \cite{pezze}: ``Not all entangled states are useful for quantum metrology''. See also \cite{bromley}. Below we demonstrate how this notion applies in the present context by analyzing the entanglement produced and the correlations present in the system.

It is convenient for our study to use the entanglement measure of {\it linear entropy}  
${\mathcal E}({\rho})$, defined  in terms of a density matrix $\rho(t)= |\Psi(t)\rangle\langle \Psi(t)|$ as \cite{zurek,buscemi} 
$
{\mathcal E}({\rho})= 1-  \rm tr{ (\rho^2)}.
$
 The linear entropy is bounded between 0 and $1-1/d$, where $d$ is the dimension of the space on which the density matrix acts. For initial state $\ket{\Psi_0} = \ket{M,P,0,0}$ the entanglement between qudits A and B  at time $t_m$ is quantified through 
$
{\mathcal E}(\rho_{1,3}(t_m)) ={1}/{2},      
$
where $\rho_{1,3}(t_m) \equiv \text{tr}_{2,4}$ $\rho(t_m)$ is the reduced density matrix (see Supplemental Material B for details). This result is independent of $P$. It asserts that immediately prior to making measurement at site 3, at time $t=t_m$, the entanglement between qudits A and B is {\it independent} of whether $N=M+P$ is even or odd.

Further, let 
$
\rho_3(t_m)={\rm tr}_1(\rho_{1,3}(t_m)),    
$
which can be compactly expressed as 
\begin{align*}
\rho_{3}(t_m)=\sum_{q=0}^M {\mathcal P}(q) \ket{q}\bra{q},    
\end{align*}
where $\mathcal{P}(q)$ refers to the probability of measuring $q$ particles at site ``3''. The linear entropy of $\rho_3$ quantifies the entanglement between the subsystems, sites 1 and 3, within qudit A.  Now we encounter a significant difference between the even and odd cases. When $N$ is odd, 
$
{\mathcal E}(\rho_3(t_m))= {1}/{2}$.     
For even $N$, $\displaystyle {\mathcal E}(\rho_3(t_m)) = 1- \frac{1}{2^{2M}} { 2M \choose M}\sim 1- \frac{1}{\sqrt{M\pi}}$,
where the second step invokes Stirling's approximation. By symmetry, the same conclusion can be drawn for qudit B (with $M$ replaced by $P$). The curious observation to make here is that in the odd case, which enables a protocol for NOON state production, the pre-measurement entanglement {\it within} the qudits is substantially {\it less} than that for the even case. This is despite the pre-measurement entanglement {\it between} the qudits being independent of number parity.  
While number-parity effects are ubiquitous in fermionic systems 
\cite{matv,zurn, matsuo,schill,mannila}, they are less frequently encountered in bosonic models. The situation reported here displays some features in common with the work of \cite{davids}.

\begin{figure}[tb]
    \centering
        \includegraphics[width=7.3cm]{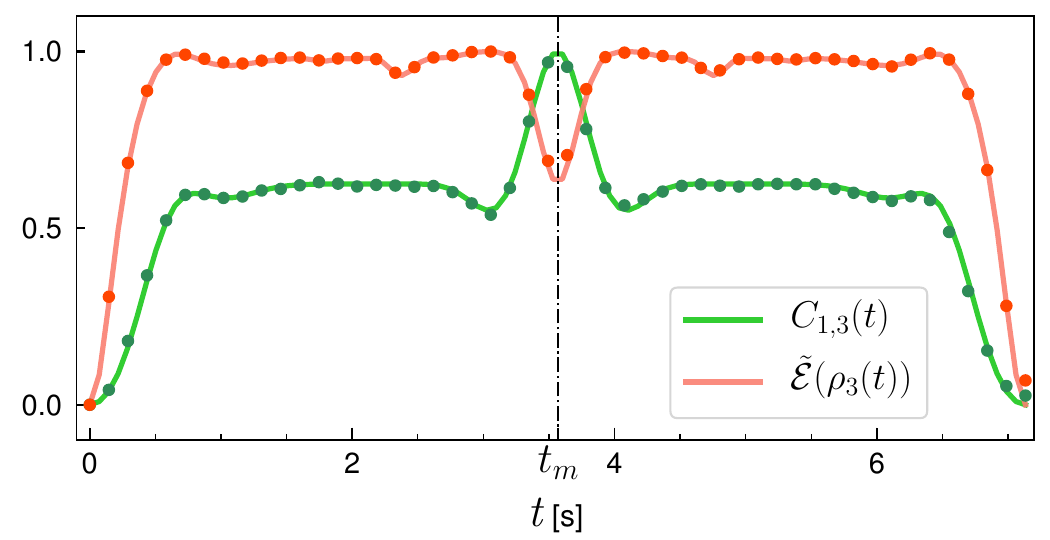}\;\;\;
    \caption{Normalized linear entropy and NOON correlation function. The red (green) line depicts $\tilde{\mathcal{E}}(\rho_3(t))$ ($C_{1,3}(t)$) calculated with the effective Hamiltonian $H_{\text{eff}}$ of (\ref{hameff}), while the dots illustrate the numerical values obtained with the Hamiltonian (\ref{ham}). The initial state is $\ket{\Psi_0} = \ket{4,11,0,0}$, and the Hamiltonian parameters are $U/\hbar = 2\pi \times 19.5$ Hz and $J/\hbar = 2\pi \times 16.2$ Hz.}
    \label{fig:entropy_corr}
\end{figure}

A similar feature is observed in the correlations of the system. In order to quantify the effects of odd/even $N$, we first define the following NOON correlation function between sites ``1'' and ``3'',
\begin{equation} C_{1,3} = \frac{4}{M^2}\left(\braket{N_1}\braket{N_3}-    \braket{N_1N_3} \right),\label{eq:correlation}\end{equation} 
where $C_{1,3}=1$ if there exists a NOON state at qudit A. 
Again for initial state $\ket{\Psi_0} = \ket{M,P,0,0}$,  using (\ref{imb1}) and the result $\langle (N_1-N_3)^2\rangle/M^2 = 1+\alpha_M\, (\mathcal{I}(2t)-1)$, $\alpha_M \equiv (M-1)/(2M)$, yields 

\begin{flalign*}
\begin{split}
C_{1,3}(t)  &= 1-\mathcal{I}^2(t)+ \alpha_M\,(\mathcal{I}(2t)-1) 
\end{split}
\end{flalign*}
and $C_{2,4}(t)=C_{1,3}(t)\Big|_{M\leftrightarrow P}$ by symmetry. At $t = t_m$, we obtain $C_{1,3}(t_m) = M^{-1}$, $C_{2,4}(t_m) = P^{-1}$ for $N$ even, and $C_{1,3}(t_m) = C_{2,4}(t_m)= 1$ for $N$ odd where the last result asserts the simultaneous existence of NOON states in each of the qudits only for the odd case. The presence of a NOON state at $t=t_m$ is signalled by attaining the maximum of the NOON correlation function $C_{1,3}$
and a simultaneous dip in the normalized linear entropy $\tilde{\mathcal{E}}(\rho_3(t))=(M+1)\mathcal{E}(\rho_3(t))/M$, as shown in Fig. \ref{fig:entropy_corr}. Further details on correlations between the qudits, and in particular the role of Eq. (\ref{imb2}), are discussed in Supplemental Material C.

\begin{figure}[t] 
    \begin{center}
         \includegraphics[width=7.9cm]{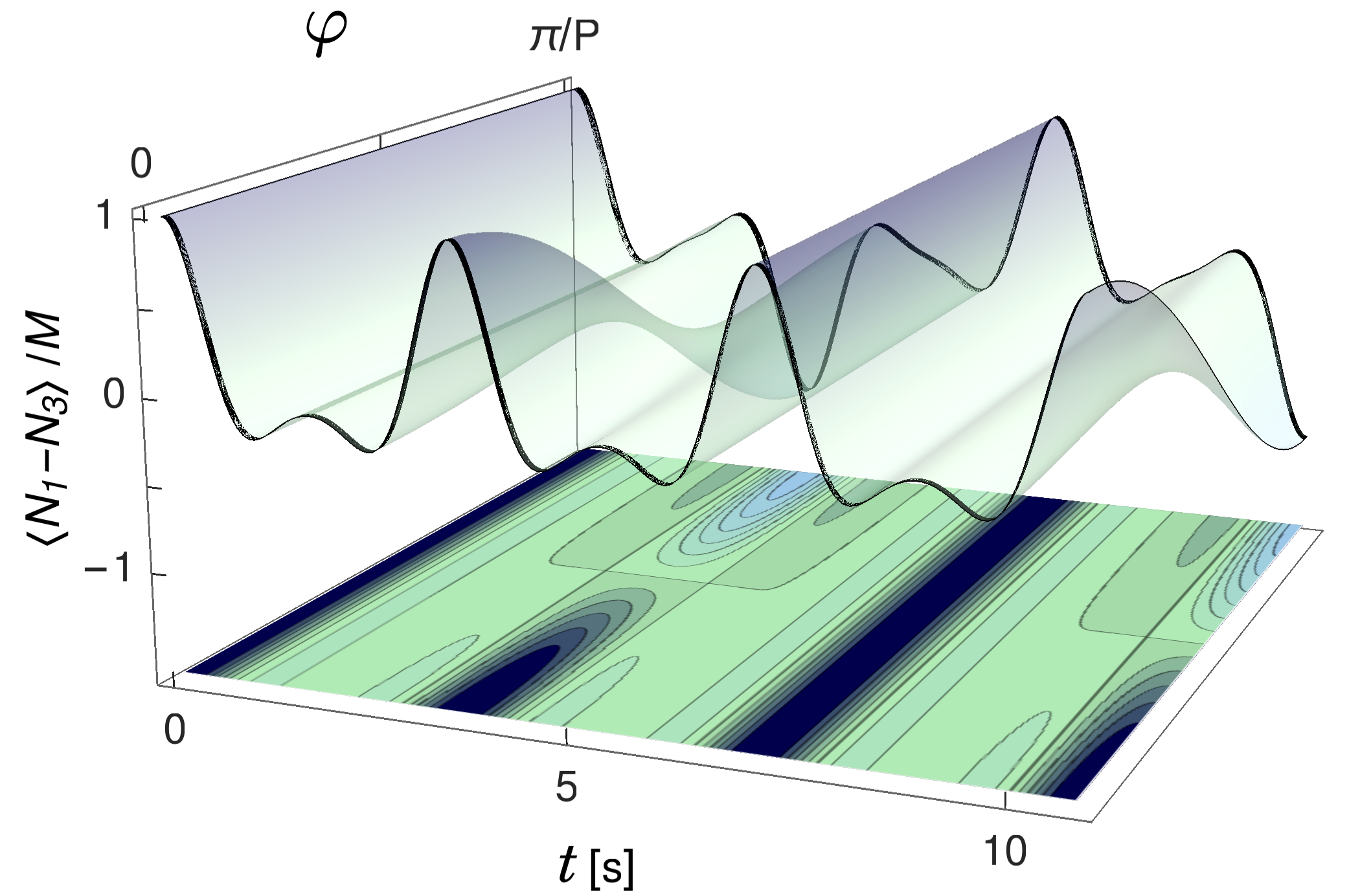} 
    \caption{Dependence of $\braket{N_1 - N_3}/M$ as a function of time $t$ (in units of seconds) and phase $\varphi$, for initial state  (\ref{noon}) with $M=4$, $P=11$, $\phi=P\varphi$, and $U/J \simeq 1.2$. Upper surface: The colors range from light to dark blue, indicating lower and higher values for the imbalance fraction. The green color represents the region where $\braket{N_1} \approx \braket{N_3}$. Lower plane: The effect on the system's dynamics is highlighted, specifically for the limiting cases $\varphi = 0$ and $\varphi = \pi/P$, where it is seen that there is a minimum-maximum inversion at $\varphi=\pi/(2P).$ }
    \label{fig:phase}
    \end{center}
\end{figure}

\textit{Heisenberg-limited interferometry.}-- 
Finally, we  establish that the system is capable of interferometry with sensitivity at the Heisenberg limit, through the archetypal example of parameter estimation through the phase of a NOON state \cite{pezze,rosetta}.   
Consider initial state (\ref{noon}) with $N=M+P$ odd, and $\phi=0$. A new phase $\varphi$ is encoded into the bosons at site 4 through a transformation, $a_4^\dagger \mapsto \exp(i\varphi) a_4^\dagger$ (cf. \cite{santos,gwylf}). This still corresponds to (\ref{noon}), but now with $\phi= P \varphi$, a phenomenon known as {\it phase super-resolution} \cite{mitchell,resch}. Again for time interval $t=t_m$, the imbalance between sites 1 and 3 is obtained from (\ref{imb2}) as 
\begin{flalign}
& \langle N_1-N_3\rangle =(-1)^{(N+1)/2}M\cos(P\varphi)
\label{fi}
\end{flalign}
providing the interference fringe with maximal contrast. Fig. \ref{fig:phase} shows the dependence of the fractional imbalance $\langle N_1 - N_3\rangle /M$
on parameters $\varphi$ and the time $t$.

Next, it can be confirmed that $\langle (N_1-N_3)^2\rangle= M^2   $, so
\begin{align*}
\Delta \langle N_1-N_3 \rangle 
&=\sqrt{\langle (N_1-N_3)^2 \rangle  - \langle N_1-N_3 \rangle^2 } \\
&= M |\sin(P\varphi)|,   
\end{align*}
where $\Delta$ denotes the standard deviation. Using the standard estimation theory approach \cite{pezze,rosetta}, it is found that the system achieves Heisenberg-limited phase sensitivity since $$\Delta \varphi=\frac{\Delta\langle N_1-N_3 \rangle}{|{\rm d}\langle N_1-N_3\rangle/{\rm d}\varphi|}=\frac{1}{P}.$$
This is an improvement on the classical shot-noise case where $\Delta \varphi\sim 1/\sqrt{P}$ \cite{pezze,rosetta}. In Supplemental Material D we present a discussion on the robustness of the system with respect to perturbation about the integrable case \cite{suppd}.


\textit{Conclusion.}-- 
We have provided an example of integrable atomtronic interferometry, through an extended Bose-Hubbard model, with four sites arranged in a closed square. The integrable properties of the model furnished the necessary tools
to understand the dynamics of the 
system in the resonant tunneling 
regime. It allowed for the analytic calculation of dynamical expectation values and correlation functions heralding NOON state formation.
 This, in turn, informed the relevant time interval  required to implement certain measurement protocols.
%
\noindent
The probabilities for measurement outcomes \cite{suppb} were computed via the density matrix.  We demonstrated proof of principle examples that the integrable system functions as an identifier of NOON states produced by a black box processor, and as a simulator of such a processor. 

Our study highlights the quantum information connections  of the model by detailing its function as a hybrid qudit system subjected to a controlled-phase gate operation.  This description complements other qudit studies in photonic 
\cite{jin,kues,imany} and NMR \cite{izi} settings, which are attracting attention due to the promise of increasing quantum computational capacity. It is anticipated that our results, in an atomtronic framework, may be transferable to these and other contexts. Besides providing feasibility for the physical set up and identifying means to experimentally probe the correlations between the qudits \cite{suppc}, 
the proposed scheme allows for further investigations of measurement-based protocols for novel quantum technologies. It also expands prospects for studying thermalization processes in the context of integrability. 

In future research, we will undertake studies involving other states that may be useful for metrological applications, such as coherent  states and Dicke states. We will examine the evolution of these input states, and investigate the correlations and the resulting generation of entanglement.  Particular emphasis will be given to the understanding of multipartite entanglement generation, beyond the bipartite investigations reported here.   

\newpage
\noindent
D.S.G. and K.W.W. were supported by CNPq (Conselho Nacional de Desenvolvimento Cient\'{\i}fico e Tecnol\'ogico), Brazil. A.F. acknowledges support from CNPq - Edital Universal 430827/2016-4. A.F. and J.L. received funding from the Australian Research Council through Discovery Project DP200101339. We thank Bing Yang and Ricardo R. B. Correia for very helpful discussions. J.L. acknowledges the traditional owners of the land on which The University of Queensland is situated, the Turrbal and Jagera people.



\clearpage
\appendix

\setcounter{page}{1}
\setcounter{equation}{0}
\renewcommand{\theequation}{S.\arabic{equation}}
\setcounter{figure}{0}
\renewcommand\figurename{Supplemental Figure}
\renewcommand{\thesubsection}{\Alph{subsection}}
\flushbottom
\section*{Supplemental Material}


\subsection{Physical setup and parameter evaluations for the integrable system}\label{SM-physical}
We describe a physical setup for the system based on superlattices, which allows for the creation of many copies of disconnected four-well square plaquettes \cite{dai2017}. This configuration can be obtained by the overlapping of two 2D optical lattices - generated separately by laser beams with wavelengths $\lambda = 532$ nm and $2\lambda$, added to a vertical 1D lattice generated by a laser with wavelength $2\lambda$ to provide pancake-shaped trapping \cite{footnote_sm}. A scheme of the superlattice is shown in Supplemental Figure \ref{fig:superl}.

\begin{figure}[h!]
    \centering
             \includegraphics[width=8.5cm]{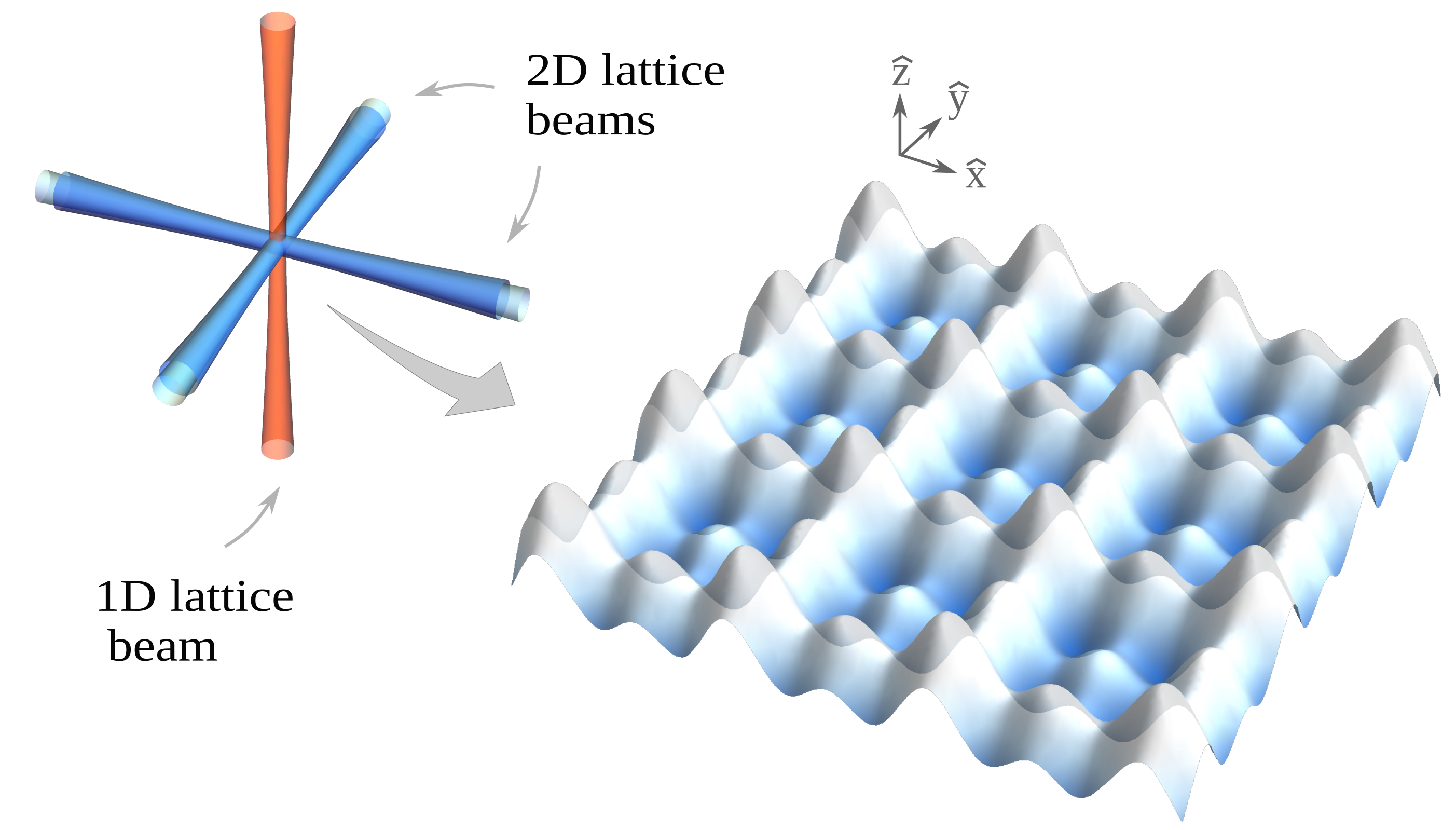}
    \caption{Schematic representation of the superlattice beams. Left, the two-dimensional lattice with two sets of counterpropagating beams crossing each other at 90$\degree$ 
    is represented in cyan. The one-dimensional vertical lattice with a counterpropagating beam along the \textbf{z}-axis, whose purpose is to create a pancake-shaped potential is represented in orange. Right, a schematic of the 2D superlattice with four-well square plaquettes.}
    \label{fig:superl}
\end{figure}

The total potential is given by:  
\begin{flalign}
\begin{split}
    V_{trap}&(x,y,z) = -V_0 \sin^2(kx) - V_0 \sin^2(ky)\\
    &+ V_1\sin^2(kx/2) + V_1\sin^2(ky/2) + \frac{1}{2}m\omega_z^2 z^2,
\label{eq:appendix-trapping-potential}
\end{split}
\end{flalign}
where $\omega_z = \sqrt{k^2 V_2 / (2m)}$ is the \textbf{z}-direction trapping frequency, $k \equiv 2\pi/\lambda$ and $m$ is the mass of the atomic species and the depths $V_i$, ($i=0,1,2$) of optical lattices control the geometry of potential trap and, consequently, the energy parameters of Hamiltonian. We set $V_1/V_0 = V_2/9V_0 = 1$, which allows for a  pancake-shaped potential trap with aspect ratio $\kappa^2\equiv\omega_z/\omega_r = 1.5$. Here we focus on only one plaquette with four sites around the origin (0,0). By performing a harmonic approximation of (\ref{eq:appendix-trapping-potential}), we obtain the potential for the $i$-th site of the chosen plaquette:
\begin{flalign}
\begin{split}
    V_{\rm trap}^{(i)}(x,y,z) =& \frac{1}{2}m\omega_r^2\left[(x-x_i)^2 
    + (y-y_i)^2\right] + \frac{1}{2}m\omega_z^2 z^2,   \nonumber
\end{split}
\end{flalign}
where $\omega_r = \sqrt{2V_0k^2/m}$ is the radial trapping frequency and $\rb_i=(x_i,y_i) = (\zeta_{i-1}d/2,\,\zeta_i d/2)$, $\zeta_i \equiv (-1)^{\lfloor i/2\rfloor}$,  is the center of site $i=1,2,3,4$. The distance between the nearest wells is given by $d=l/\delta$, where $l = \pi/k = \lambda/2$ is the usual lattice spacing constant, and $\delta = \left[1-V_1/(2\pi V_0)\right]^{-1}$ is a constant that arises in this approximation. 

We assume that the atoms are tightly confined in the trap, such that the wavefunction for each site - $\phi_i(\rb)$ - can be defined as the ground-state of $V_{\rm trap}(\rb)$ in the harmonic approximation. In the usual second quantization formalism, the bosonic field operator is written as $\Psi(\rb)=\sum_{i=1}^{4}\phi_i(\rb)a_i$.
Due to the tight confinement, the second-nearest-neighbor tunneling term vanishes, and we can recover the Hamiltonian (\ref{ham}), where the parameters are defined as $ \displaystyle U_0 =  g\int d\rb \sp |\phi_1(\rb)|^4+U_{11}$, $\displaystyle U_{ij} = \frac{C_{dd}}{4\pi}\int d\rb \int d\rbb |\phi_i(\rb)|^2|\phi_j(\rbb)|^2\frac{1-3\cos^2\theta_P}{|\rb-\rbb|^3}$ and $\displaystyle J =2 \int d\rb\sp \phi_1(\rb)\left[\frac{\hbar^2\nabla^2}{2m} - V_{trap}(\rb)\right]\phi_2(\rb)$.

The respectively on-site and inter-site energies $U_0$ and $U_{ij}$ are consequences from the interaction potential :
$$V(\rb - \rbb) = \underbrace{g\,\delta(\rb-\rbb)}_{V_{SR}} + \underbrace{\frac{C_{dd}}{4\pi}\frac{1-3\cos^2\theta_P}{|\rb-\rbb|^3}}_{V_{DDI}}.$$
The first term ($V_{SR}$) characterizes the short-range interaction with coupling constant $g = 4\pi\hbar^2 a/m$, where $a$ is the s-wave scattering length, which can be positive or negative - yielding, respectively, a repulsive and attractive contact interaction -, and whose values can be tuned through a magnetic field via Feshbach Resonance. In the dipole-dipole interaction (DDI) potential ($V_{DDI}$), $C_{dd} = \mu_0 \mu^2$, where $\mu_0 = 4\pi \times 10^{-7} N/A^2$ is the vacuum magnetic permeability, $\mu$ is the atom's magnetic dipole moment and $\theta_P$ is the angle between the dipole orientation and the direction of $\rb-\rbb$. 
The insertion of $\phi_i(\rb)$ as the ground-state of $V_{\text{trap}}(\rb)$ in the above equations
results in
$\displaystyle     U_0 =\kappa\left(\frac{\eta}{\pi}\right)^{3/2}\left(g - \frac{C_{dd}}{3}f(\kappa)\right)$
$\displaystyle     U_{ij} = \frac{C_{\text{dd}}}{4\pi}\int_{0}^{\infty}  {\rm d}r\, r\exp\left(-\frac{r^2}{4\eta}\right) J_0(r d_{ij}) Z(r)$,
%
with $ \displaystyle Z(r)= \frac{4}{3}\sqrt{\frac{\kappa^2\eta}{\pi}}  - r\exp\left( \frac{r^2}{4\kappa^2\eta}\right) \text{erfc}\left(\frac{r}{2\sqrt{\kappa^2\eta}}\right)$ and $J_0$ is the Bessel function of first kind. We define $\eta \equiv m\omega_r/(2\hbar)$, and $d_{ij}$ stands for the distance between the two sites. Here we also notice that, in the limit $d_{ij}\rightarrow0$, $\displaystyle U_{ij}\rightarrow -\frac{\kappa C_{dd}}{3}\left(\frac{\eta}{\pi}\right)^{3/2}f(\kappa)$, where $f(\kappa)$ is a function that relates the trap aspect ratio $\kappa^2$ with the DDI \cite{lahaye}. A schematic representation of the interactions $U_0$ and $U_{ij}$ is depicted in Supplemental Figure \ref{Us}.

\begin{figure}
    \centering
    \includegraphics[width=8cm]{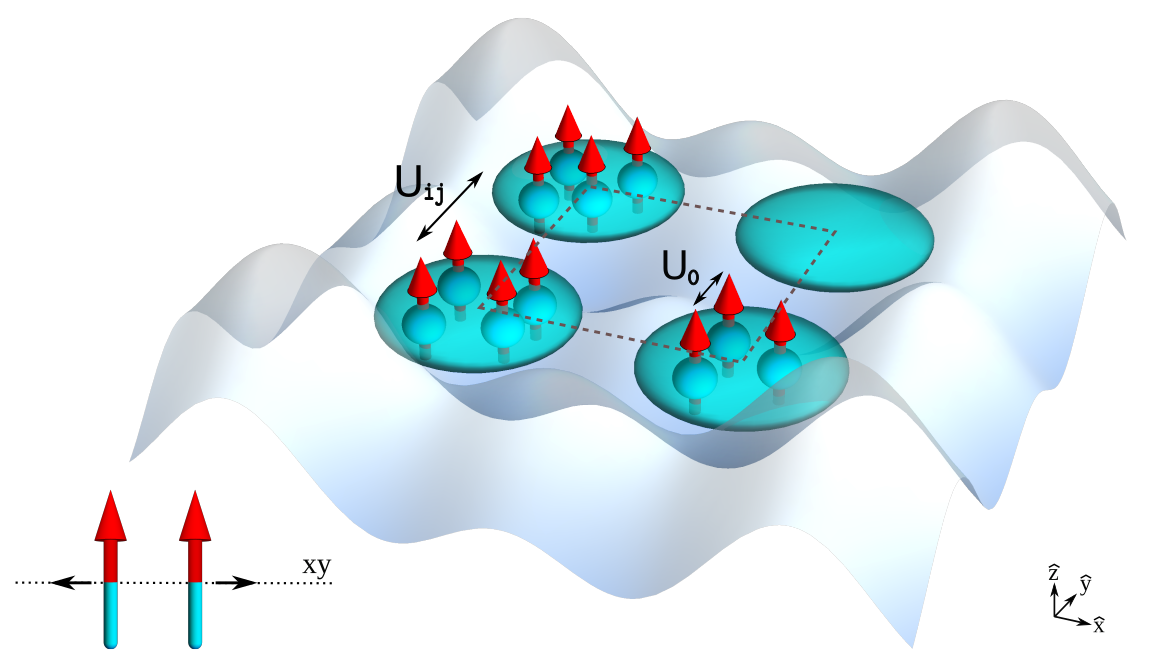}
    \caption{ Schematic representation of the on-site ($U_0$) and inter-site ($U_{ij}$) interactions of ultracold Bose gases trapped in a four-site square plaquette. The oblate shapes in cyan represent the trap and the arrows depict the polarization of the dipoles aligned in the $z$-direction. The resulting parallel alignment of the dipoles causes a repulsive dipole-dipole interaction both within the same well and between neighboring wells, as depicted in the sketch on the lower left.}
    \label{Us}
\end{figure}

As stated in the article, the Hamiltonian acquires two additional independent, conserved operators when the integrability condition, $U_0 - U_{13} = 0$, is fulfilled.
This condition can be solved numerically in the following way: by choosing a value for the s-wave scattering length ($a$), the value of the parameter $\omega_r$ is varied up to the point where $U_0$ is equal to $U_{13}$. This means that it is the condition of integrability that states the value for the trapping frequencies, thereby ensuring it's fulfilment.

For the Hamiltonian parameters in this work, we considered $^{164}$Dy, for which the magnetic dipole moment is $\mu =9.93$  (in units of Bohr magneton $\mu_B$), with $a = -21.4$ (in units of Bohr radius $a_0$) and $V_1/V_0 = V_2/9V_0 = 1$. Here, the DDI provides significant long-range interactions 
and the trap geometry favors the repulsive on-site DDI which dominates
the short-range attractive contact interaction, such that $U_0>0$. From the integrability condition, we obtain $\omega_r \approx 2\pi\times51.46$ kHz, 
resulting in 
$U_0/\hbar \approx 2\pi\times41.38$ Hz, $U_{12}/\hbar \approx 2\pi\times119.25$ Hz 
and $J/\hbar \approx 2\pi\times16.18$ Hz, as used throughout the analysis.


\subsection{{Probabilities, fidelities and trade-off  between  fidelity  and  protocol  time}}\label{SM-prob-fid}
Here we provide benchmarks establishing the effectiveness of  Hamiltonian (\ref{ham}) in the simulation of the black box processor ${\mathcal P}$, through numerical calculation of probabilities and outcome fidelities.
A general $N$-particle state is expressed as 
$ \displaystyle
|\Theta\rangle= \sum _{ j,k,l=0 }^N  c_{j,k,l}   |j,k,l, N-j-k-l\rangle
$ 
such that $c_{j,k,l}=0$ if $j+k+l> N$, and
$\displaystyle
\sum_{j,k,l=0}^N |c_{j,k,l}|^2=1$. 
When a measurement is made at site 3, the probability ${\mathcal P}(r)$ to obtain the measurement outcome $r$ is 
\begin{align}
{\mathcal P}(r) = \sum_{j,k=0}^N |c_{j,k,r}|^2
\label{prob}
\end{align}
satisfying
$\displaystyle 
\sum_{r=0}^N {\mathcal P}(r)=1$.
After the measurement, the wavefunction collapses to 
\begin{align*}
|\Theta(r)\rangle= \frac{1}{\sqrt{{\mathcal P}(r)}} 
\sum _{ j,k=0}^N  c_{j,k,r}   |j,k,r, N-j-k-r\rangle
\end{align*}
 such that $\langle \Theta(r) |\Theta(r)\rangle=1$.
Set 
\begin{align*}
\ket{\Phi(r,\phi)} = \frac{1}{\sqrt{2}} \ket{M-r,P,r,0} + \frac{\exp{(i\phi)}}{\sqrt{2}}\ket{M-r,0,r,P}
\end{align*}
and define the outcome fidelity ${\mathcal F}(r,\phi)$ to be 
\begin{align}
{\mathcal F}(r,\phi)&= | \langle \Phi(r,\phi)|\Theta(r)\rangle |.
\label{fid}
\end{align}
We take 
$  
\ket{\Theta} = \exp(-it_m H)\ket{4,11,0,0}    
$ 
and use (\ref{ham}) with $U/\hbar \simeq 2\pi\times19.5$ Hz, $J/\hbar \simeq 2\pi\times16.2$ Hz to numerically calculate the measurement probabilities and outcome fidelities through (\ref{prob},\ref{fid}). Results are given in Table I. 
\begin{table}[h]
\begin{center}
\def\arraystretch{1}
\resizebox{5.2cm}{!}{
\begin{tabular}{cccc}
\hline
Measurement & Probability  & Phase  & Fidelity  \\
$r$ & ${\mathcal P}(r)$ & $\phi$ & ${\mathcal F}(r,\phi)$ \\
\hline 
 4 & $0.47639$ & $\pi$ & $0.99298$  \\
 3 & $0.00729$ & $\pi$ & $0.12352$  \\
 2 & $0.00368$ & $\pi$ & $0.02552$  \\
 2 & $0.00368$ & $0$ & $0.01375$  \\
 1 & $0.00625$ & $0$ & $0.13710$  \\
 0 & $0.49611$ & $0$ & $0.97831$  \\
\hline
\end{tabular}}
\caption{Measurement probabilities and fidelities after evolution under (\ref{ham}) until time $t_m$. The initial state is $\ket{4,11,0,0}$, and $U/\hbar \simeq 2\pi\times19.5$ Hz, $J/\hbar \simeq2\pi\times16.2$ Hz as used in Figs. \ref{fig:two}, \ref{fig:phase} of the main text.
The calculations show that the highest fidelity outcomes, close to 1, occur with the highest probabilities, close to 1/2. This is in  agreement with the results predicted by the effective Hamiltonian (\ref{hameff}). }
\label{table:1}
\end{center}
\end{table}

To analyze the trade-off between protocol time and fidelity, we evaluate the fidelity \eqref{fid} for different values of $t_m$ (determined by different values of $U$) for both $\phi =0$ and $\pi$. We find that the system reaches proximity to the resonant regime for parameters $U$ and $J$ such that $t_m > 1.75$s. Note that lifetimes of atoms in optical lattices may be as large as 300s \cite{gibbons}. As seen in Supplemental Figure \ref{fig:fid}, the parameters $U/\hbar \simeq 2\pi\times19.5$ Hz and $t_m \simeq 3.57$s (used in the main text and marked by the dashed vertical line) are within the stable resonance region.
\begin{figure}[h]
    \centering
    \includegraphics[width=7.5cm]{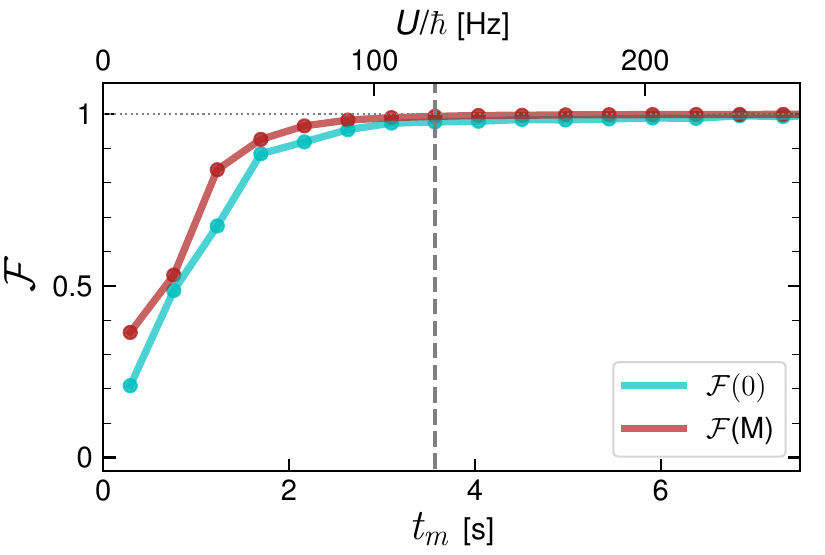}
    \caption{Trade-off between fidelity and protocol time. The fidelities of the NOON states (obtained after the measurement procedure) for initial state $|4,11,0,0\rangle$ with $J/\hbar\simeq 2\pi\times16.2$ Hz and varying $U/\hbar$ from $\sim2\pi\times2$ Hz to $\sim2\pi\times40$ Hz (top axis), such that $t_m$ varies accordingly between $\sim0.3$ s and $\sim7.3$ s. Cyan line: $\phi=0$. Brown line: $\phi=\pi$. The vertical dashed line shows $t_m \simeq 3.57$s corresponding to $U/\hbar \simeq 2\pi\times19.5$ Hz as used in the main text. }
    \label{fig:fid}
\end{figure}

%
One of the counter-intuitive features of this theoretical framework is the acute dependence on whether the total particle number $N=M+P$ is even or odd. To provide an understanding of this phenomenon, we take the initial state $\ket{M,P,0,0}$ and consider the time evolution of the {\it reduced density matrix} $\rho_{1,3}(t)$ for qudit A, 
$\rho_{1,3}(t) = {\rm tr}_{2,4}\left( |\Psi(t)\rangle\langle \Psi(t)| \right)$, 
where ${\rm tr}_{2,4}$ is the partial trace taken over the state space for qudit B, and 
$|\Psi(t)\rangle= \exp(-i t H_{\rm eff}) \ket{M,P,0,0}$.
We then obtain
\begin{align*}
\rho_{1,3}\left(t_m\right)  
&=\frac{1}{2} |\Psi_+\rangle\langle \Psi_+| +\frac{1}{2}|\Psi_-\rangle \langle \Psi_- | , \\
    \ket{\Psi_\pm} &= \frac{1}{\sqrt{2^{M}}}  \sum_{r=0}^{M} \sqrt{M\choose r } \exp\left(-i\frac{(N\pm 1)r\pi }{2}\right) \ket{\chi(r)}, \\
\ket{\chi(r)}&=\frac{1}{\sqrt{2^M(M-r)!r!} }   
    (a_1^\dagger+a_3^\dagger )^{M-r} 
    (a_1^\dagger - a_3^\dagger )^r\ket{0}.  
\end{align*}
The above results then allow for a calculation of the probability ${\mathcal P}(r)$ that,  measurement of the number of particles at site 3, when $t=t_m$, yields the outcome $r$, resulting
\begin{small}
\begin{align*}
{\mathcal P(r)}&= \frac{1}{2}b_{M,r}(\sin^2((N-1)\pi/4))+
\frac{1}{2}b_{M,r}(\sin^2((N+1)\pi/4))
\end{align*}
\end{small}
\noindent where
$\displaystyle 
b_{M,r}(x)= {M \choose r}   x^r(1-x)^{M-r},
\, r=1,...,M,
$ 
are the {\it Bernstein polynomials}. When $N$ is even, 
$ \displaystyle 
{\mathcal P}(r)=\frac{1}{2^M} {M\choose r}.
$ 
When $N$ is odd, ${\mathcal P}(r)=(\delta_{r,0}+\delta_{r,M})/2$.
The binomial distribution of the even case has maximal support, in stark contrast to the double-delta function distribution of the odd case. 



\subsection{Correlations between qudits}\label{SM-correl}
 The interdependence between the measurement of particle number in qudit A and the symmetry of NOON state produced in qudit B,  as discussed below Eq. (\ref{gennoon}), can be characterized in terms of the {\it swapped fractional imbalance correlation} (SFIC). This is defined as $C_{AB}^{X} \equiv \langle I_A V_B\rangle-\langle I_A\rangle \langle V_B\rangle$ for a given initial state $|X\rangle$, where $I_A=(N_1-N_3)/M$ is the fractional imbalance operator of the qudit A and $V_B$ is the {\it swap operator} \cite{islam} acting on qudit B as $V_B|a,b,c,d\rangle = |a,d,c,b\rangle$. Since $V_B^2 = Id$, the swap operator has eigenvalues $\pm 1$ that allow to distinguish between a symmetric ($+1$) or antisymmetric ($-1$) NOON state in qudit B. The value of the SFIC is bounded, $|C_{AB}^{X}|\leq 1$. After the initial state $|\Psi_0\rangle =|M,P,0,0\rangle$ evolves for time $t=t_m$, the SFIC achieves the boundary values $C_{AB}^{\Psi_0} = +1$ or $C_{AB}^{\Psi_0}=-1$. 
Throughout the evolution from the initial state $|\Psi_0\rangle$ the expectation value of the swap operator is $\langle V_B\rangle = 0$. 
This result allows Eq. (\ref{imb2}) to be formulated as 
\begin{align}
\mathcal{I}_{\Phi(\phi)}(t)&\equiv \langle \Phi(\phi)|e^{iHt}I_Ae^{-iHt}|\Phi(\phi)\rangle \nonumber \\
&= \mathcal{I}(t)+C_{AB}^{\Psi_0}\cos\phi,
\label{re}
\end{align} 
and the SFIC for initial NOON state $|\Phi(\phi)\rangle$ as a phase-controlled quantity given by $C_{AB}^{\Phi} = C_{AB}^{\Psi_0}\sin^2\phi$. These formulae reveal two important aspects: ($i$) since $|C_{AB}^{\Psi_0}| = 1$ at $t=t_m$, the interference fringe emerges with maximal contrast when  $|C_{AB}^{\Phi}|$ achieves its maximum, see Eq. (\ref{fi}); ($ii$) Eq. (\ref{re})  provides $C_{AB}^{\Psi_0}=\mathcal{I}_{\Phi(0)}(t)-\mathcal{I}(t)$, meaning the SFIC $C_{AB}^{\Psi_0}$ can be determined in terms of fractional imbalances for the initial states $|M,P,0,0\rangle$ and the NOON state $|\Phi(0)\rangle$. The SFIC $C_{BA}^{\Psi_0}$ can be defined in an analogous way. These correlations simultaneously attain the maximum value $|C_{BA}^{\Psi_0}| = |C_{AB}^{\Psi_0}| = 1$ at $t=t_m$. 
 
\subsection{Robustness of the interferometer}\label{SM-robust}

An important consideration is the robustness of an integrable system against noise \cite{tang2018}. 
Here, we discuss the interferometric performance under perturbations away from integrability,  allowing for $U_0-U_{13} \neq 0$. Specifically,
$$ 
H_{\epsilon} = H +\epsilon\nu (N_1 N_3+N_2 N_4).
$$ 
where the parameter $\epsilon$ has the magnitude of $U_0-U_{13} \neq 0$ and $\nu\in[-1,1]$ is a random fluctuation number that models experimental noise, such as that resulting from laser intensity variations or magnetic field fluctuations \cite{oberthaler2010}.

To quantify the robustness, we examine the time evolution of the fidelity between the initial state  
\begin{align*}
|\Psi\rangle=\frac{1}{\sqrt{2}}(|4,11,0,0\rangle+|4,0,0,11\rangle),
\end{align*} 
evolving under $H$, and the same state evolving under $H_{\epsilon}$. The state evolving under $H$ is expressed as 
$
|\Psi(t,0)\rangle = \exp(-i t H)|\Psi\rangle    
$
while the state evolving under $H_\epsilon$ is iteratively defined as
$ \
|\Psi(t+\delta t,\epsilon)\rangle = \exp(-i \delta t H_\epsilon)|\Psi(t,\epsilon)\rangle    
$ 
where the fluctuation $\nu$ is applied in each time step $\delta t = t_m/\mathcal{N} \sim$ 4 ms. 
Then for each value of $\epsilon$,  100 numerical data points are considered to compute the average fidelity
$ 
F_{\epsilon}(t)= \overline{|\langle\Psi(t,0)|\Psi(t,\epsilon)\rangle|}$.
The results are presented in Supplemental Figure \ref{fid_epsilon}.
    \begin{figure}[h]
    \centering
        \includegraphics[width=1.\linewidth]{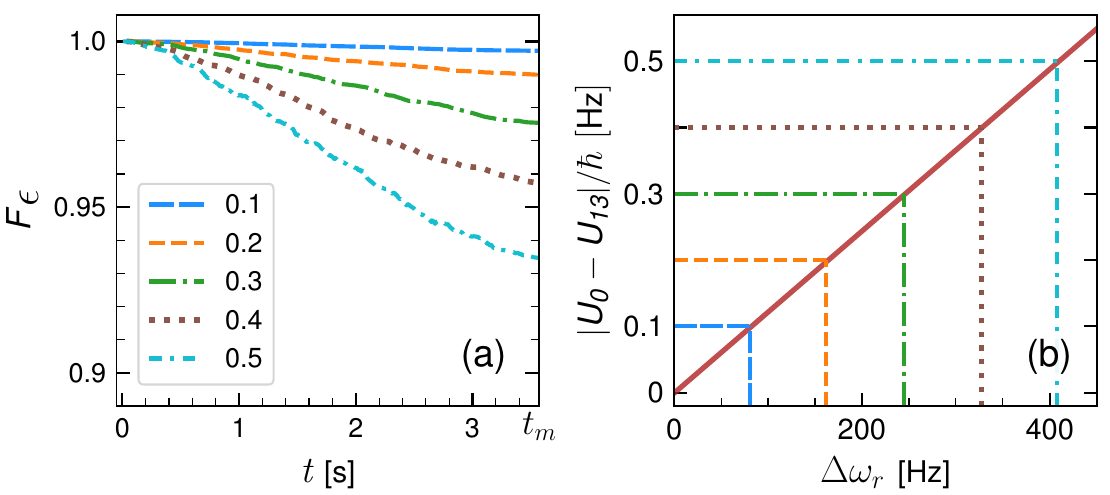}
        \caption{Robustness analysis. {\bf (a)} time evolution of the fidelity $F_{\epsilon}$ until $t=t_m \simeq 3.57$s (endpoint of the axis), for different values of the parameter $\epsilon =0.1,\,0.2,\,0.3,\,0.4,\,0.5$ (in units of $\hbar\,$s$^{-1}$, according to the legend) with $\mathcal{N}=900$, $\phi = 0$, $U/\hbar \simeq 2\pi\times19.5$ Hz and $J/\hbar \simeq 2\pi\times16.2$ Hz.
       {\bf (b)} $|U_0-U_{13}|$ vs $\Delta \omega_r = |\omega_r'-\omega_r|$  where $\omega_r = 2\pi\times 51.4619$ kHz at $U_0=U_{13}$. The dashed lines mark the cases where $|U_0-U_{13}| = 0.1,\,0.2,\,0.3,\,0.4,\,0.5$ (in units of $\hbar\,$s$^{-1}$). 
        }
        \label{fid_epsilon}
    \end{figure}
    

\renewcommand{\refname}{\bf Supplemental Material References}



\begin{thebibliography}{99}
\bibitem{footnote_sm} A setup for cigar-shaped trapping is discussed in \cite{noon}.
\end{thebibliography}
\end{document}